\documentclass[epj]{webofc}

\begin{document}

\title{\textbf{Integrable cosmological  models with non-minimal  coupling  and bounce solutions}}
\author{\firstname{Ekaterina} \lastname{Pozdeeva}\inst{1}\fnsep\thanks{\email{pozdeeva@www-hep.sinp.msu.ru}} \and
        \firstname{Sergey} \lastname{Vernov}\inst{1}\fnsep\thanks{\email{svernov@theory.sinp.msu.ru}}
}

\institute{Skobeltsyn Institute of Nuclear Physics,  Lomonosov Moscow State University, Leninskie Gory 1, 119991, Moscow, Russia }

\abstract{
  We remind the way to obtain integrable system with non-minimally coupled scalar fields. We are interesting to models with bounce solutions and compare bounce solutions in for two known integrable models. We show that only one model has a bounce solution that tends to a stable de Sitter solution.
}
\maketitle



\section{Introduction}

Models with scalar fields  are useful to describe the observable evolution of the Universe as the dynamics of the spatially flat FLRW background with the interval
\begin{equation}
\label{Fried}
ds^2=N^2(\tau) d\tau^2-a^2(\tau)\left(dx_1^2+dx_2^2+dx_3^2\,\right),
\end{equation}
where $a(\tau)$ is the scale factor, $N(\tau)$ is the lapse function of the parametric time $\tau$,
and cosmological perturbations.

The recent high-precision measurements by the Planck telescope~\cite{Planck2015}  confirm predictions of the single-field inflationary models. At the same time, the predictions of the simplest inflationary models with a minimally coupled scalar field lead to sufficiently large values of the tensor-to-scalar ratio of the density perturbations $r$, and are, therefore, ruled out by Planck data.
Such type of inflationary scenarios has the singularity problem. Indeed, the Hawking--Penrose singularity theorems prove that the Universe in models with standard scalar fields minimally coupled to gravity is geodesically past incomplete. At the same time bouncing cosmological scenario with non-minimally coupled scalar fields naturally avoids this singularity problem.

At the bounce point the period of universe contraction changes to a period of universe expansion. Thereby, a bounce point is characterized by two condition: at this point the Hubble parameter is equal to zero and its cosmic time derivative is positive.
 Note that in models with standard (not phantom) scalar fields minimally coupled yo gravity the Hubble parameter is monotonically decreasing function. The simplest way to get non-monotonic behavior of the Hubble parameter is to introduce in the model both standard and phantom scalar fields and consider so-called quintom models~\cite{QUINTOM,AKV2}.
In phantom field models the Null Energy Condition is violate and instability problems arise~\cite{Rubakov:2014jja}. To avoid these problem bouncing models with Galileon fields have been constructed~\cite{Qiu:2011cy,GBounce,Osipov:2013ssa,Koehn:2013upa,Libanov:2016kfc,Ijjas:2016tpn}. Another possibility to get non-motonic behavior of the Hubble parameter  is  to consider models with the Ricci scalar multiplied by a function of the standard scalar field~\cite{Tretyakova2011,KTVV2013,Pozdeeva2014,Boisseau:2015hqa,KPTVV2015,Boisseau:2016pfh,PSTV2016}.

In spite of great success of numerical simulations and different approximation schemes for studying of cosmological models, the exact solutions are always useful and allow to catch some qualitative features of differential equations.
The use of the FLRW metric essentially simplify the Einstein equations. But, only a few cosmological models with minimally coupled scalar fields  are integrable~\cite{SalopekBond,Marek,BarsChen,ACK,Fre,Fre2,Boisseau:2015cda}.
The list of such models has been presented in~\cite{Fre}. In~\cite{KPTVV2013}  a method for constructing integrable models with non-minimally coupled scalar fields by using the interrelation between the Jordan and Einstein frames has been proposed.  Sometimes the integrability of the non-minimally coupled model is more apparent that the integrability of its minimally coupled counterpart~\cite{Borowiec:2014wva,Boisseau:2015hqa}. So it is useful to connect integrable models in the Einstein frame and the Jordan frame~\cite{KPTVV2013,Boisseau:2015cda,KPTVV2015}.

In~\cite{Boisseau:2015hqa}, a flat FLRW cosmological model having  the Hilbert--Einstein term, a positive cosmological constant, and a conformally coupled scalar field with a negative quartic potential was investigated.  Such a model is exactly integrable and for a large class of initial conditions possesses a bounce, so, it avoids
the cosmological singularity. The key property of this model is that the  Ricci scalar $R$ is an integral of motion. In~\cite{KPTVV2015} it has been shown that this model with a constant $R$ belongs to one-parameter set of integrable models, moreover, some of these integrable models have bounce solutions and polynomial potentials as well. In this short paper we compare the bouncing solutions obtained in different integrable models with non-minimal coupled scalar fields.

\section{The simplest integrable model with bounce solutions}

Let us consider a cosmological model, described by the following action
\begin{equation}
S =\int d^4x\sqrt{-g}\left[U(\varphi)R - \frac12g^{\mu\nu}\varphi_{,\mu}\varphi_{,\nu}-V(\varphi)\right],
\label{action}
\end{equation}
where $U(\varphi)$ and $V(\varphi)$ are differentiable functions of the scalar field $\varphi$, $g=\text{det}(g_{ik})$ is the determinant of the metric tensor $g_{ik}$, $R$ is the Ricci scalar.
Varying  action (\ref{action}) and substituting the FLRW metric~(\ref{Fried}), one can obtain the following equations~\cite{Kamenshchik:1998ue,KPTVV2015}:
\begin{equation}
\label{Frequoc00}
6Uh^2+6U'h\dot{\varphi}=\frac12\dot{\varphi}^2+N^2V,
\end{equation}
\begin{equation}
\label{Frequocii}
 2U\left[2\dot{h}+3h^2-2h
 \frac{\dot{N}}{N}\right]
+2U'\left[\ddot{\varphi}+2h\dot{\varphi}-\frac{\dot{N}}{N}\dot{\varphi}\right] =N^2V-\left[2U''+\frac12\right]
\dot{\varphi}^2,
\end{equation}
\begin{equation}
\ddot{\varphi}+\left(3h-\frac{\dot{N}}{N}\right)\dot{\varphi} -6U'\left[\dot{h}+2h^2-h\frac{\dot{N}}{N}\right]+N^2V' = 0,
\label{KGoc}
\end{equation}
where the function $h=\dot{a}/a$, a ``dot'' means a derivative with respect to the parametric time $\tau$ and a ``prime'' means a derivative with respect to the scalar field.
If $N(\tau)\equiv1$, then $h(\tau)$ is the Hubble parameter denoted as $H(t)$, where $t=\tau$ is the cosmic time.

If $U=U_0$ is a constant, then the scalar field minimally coupled to gravity and system (\ref{Frequoc00})--(\ref{KGoc}) looks simpler, but even in this case the standard way to prove the integrability includes the suitable choice of the function $N(\tau)$ that allows to simplify or even to linearize the equations~\cite{Fre}. At the same time there exists the integrable system with non-minimally coupled scalar field that integrability is obvious and one do not need to guess~$N(\tau)$.

As it has been shown in~\cite{PSTV2016} from
system (\ref{Frequoc00})--(\ref{KGoc}) with $N=1$ one can obtain the following equation
\begin{equation}
\label{equR}
2R\left( U+3{U'}^2 \right) +\left(6 U''+1\right) {\dot {\varphi}}^{2}=4V
+6 V'U'\,,
\end{equation}
where the Ricci scalar $R=6(\dot H+2H^2)$.

It is easy to see that a constant $R$ is a solution of this equation if
\begin{equation}
\label{Uc}
   U=U_c=U_0-\frac{1}{12}\varphi^2,\qquad V=V_{c}=C_0+C_4\varphi^4,
\end{equation}
where $U_0$, $C_0$ and $C_4$  are constants.  Indeed, substituting $U_c$ and $V_c$ into Eq.~(\ref{equR}),  we get
\begin{equation}
\label{Rconst}
R =2\frac{C_0}{U_0}.
\end{equation}

Formula~(\ref{Rconst}) defines a differential equation for the Hubble parameter:
\begin{equation}
\label{equHR}
3\left(\dot H+2H^2\right)=\frac{C_0}{U_0},\qquad \Leftrightarrow \qquad
3\left(\ddot a a+\dot a^2\right)=\frac{C_0}{U_0} a^2.
\end{equation}
Therefore, there exists the following integral of motion~\cite{PSTV2016}:
\begin{equation*}
I=3\dot{a}^2 a^2-\frac{C_0}{4U_0} a^4.
\end{equation*}

The considering integrable  model is interesting due to bounce solutions.
Let us remind that a bounce point $t_b$ is defined by two conditions: the Hubble parameter $H(t_b)=0$ and $\dot{H}(t_b)>0$.
From system (\ref{Frequoc00})--(\ref{KGoc})  with $N=1$ and $U=U_c$ , one gets the following condition of the potential:
\begin{equation}
V(\varphi(t_b))<0, \qquad 4V(\varphi(t_b))-\varphi(t_b) V'(\varphi(t_b)) > 0.
\label{bouncecond}
\end{equation}

For $V=V_{c}$ from~(\ref{bouncecond}), we get $C_0>0$ and $C_4<0$.
Equation~(\ref{equHR}) with a positive $C_0$ has two possible real solutions in dependence of the initial conditions:
\begin{equation}
\label{Hmonot}
H_1(t) = \sqrt{\frac{C_0}{6U_0}}\tanh\left(\sqrt{\frac{2C_0}{3U_0}}(t-t_0)\right),\qquad H_2(t) = \sqrt{\frac{C_0}{6U_0}}\coth\left(\sqrt{\frac{2C_0}{3U_0}}(t-t_0)\right),
\end{equation}
where $t_0$ is an integration constant. To get a standard gravity domain with $U>0$ we assume $U_0>0$. The behavior of the Hubble parameter does not depend on the specific dynamics of the scalar field $\varphi$, because two-parametric set of functions $\varphi(t)$  corresponds to one-parametric set of $H(t)$. At the same time not all solutions with $H_1(t)$
tend to de Sitter ones, because the scalar field $\varphi$ may tend to infinity, hence, the function $U(\varphi)$ stands negative.

System (\ref{Frequoc00})--(\ref{KGoc}) with $N=1$  and $U=U_c$ can be transformed into the following dynamical system~\cite{KPTVV2016}:
\begin{equation}
\label{FOSEQU}
\left\{
\begin{split}
\dot\varphi &=\psi,\\
\dot\psi&={}-3H\psi-\frac{\left(12U_0-\varphi^2\right)V'+4\varphi V}{12U_0},\\
\dot H&={}-\frac{1}{12U_0}\left(2\varphi^2H^2+\left[4H\psi-V'\,\right]\varphi+2\psi^2\right).
\end{split}
\right.
\end{equation}

Equation~(\ref{Frequoc00})  with $N=1$  and $U=U_c$  has the following form

\begin{equation}
\label{Frequoc00N1}
6H^2\left(U_0-\frac{1}{12}\varphi^2\right)- H\varphi\dot{\varphi}-\frac12\dot{\varphi}^2-V=0\,.
\end{equation}
If Eq.~(\ref{Frequoc00N1}) is satisfied in the initial moment of time, then it is satisfied at any moment of time. By this reason Eq.~(\ref{Frequoc00N1}) fixes initial conditions of system~(\ref{FOSEQU}).

To analyze qualitative behaviour of solutions, in particular, to analyze  the stability of de Sitter solutions it is useful to introduce new variables~\cite{Skugoreva:2014gka}, namely, the effective potential
\begin{equation}
\label{Veff}
V_{eff}(\varphi)=\frac{U_0^2V(\varphi)}{U(\varphi) ^2}.
\end{equation}
and  functions
\begin{equation}
\label{PA}
P\equiv \frac{H}{\sqrt{U}}+\frac{U'\dot\varphi}{2U\sqrt{U}},\qquad A\equiv\frac{U+3{U'}^2}{4U^3}.
\end{equation}

From Eqs.~(\ref{Frequoc00}) and (\ref{Frequocii}) we get equations that look like the Friedmann equations for models with  minimally coupling:
\begin{equation}
\label{equP}
3P^2=A{\dot\varphi}^2+\frac{1}{2U_0^2}V_{eff},\qquad
\dot P={}-A\sqrt{U}\,{\dot\varphi}^2.
\end{equation}
If $U(\varphi)>0$, then $A(\varphi)>0$ as well. Therefore, the function $P$ is a monotonically decreasing function at $U>0$.

De Sitter solutions correspond to extrema of the effective potential:  $V'_{eff}(\varphi_{dS})=0$. If $U>0$, then the model has a stable de Sitter solution only if $V_{eff}(\varphi_{dS})>0$ and $V''_{eff}(\varphi_{dS})>0$~\cite{PSTV2016}. Let us clarify the condition $V_{eff}(\varphi_{dS})>0$ . This condition is equivalent to $V(\varphi_{dS})>0$. From Eq.~(\ref{Frequoc00}) we obtain
\begin{equation}
\label{Hf}
H_{dS}=\pm\sqrt{\frac{V(\varphi_{dS})}{6U(\varphi_{dS})}}.
\end{equation}
If $U(\varphi_{dS})>0$, then $H_{dS}$ is a  nonzero real number only at $V(\varphi_{dS})>0$. In the antigravity region with $U<0$ one has the following conditions for a stable de Sitter solution: $V_{eff}(\varphi_{dS})<0$ and $V''_{eff}(\varphi_{dS})<0$.

If the function $U$ is always positive, then one can use the equations in terms of $P$, $A$, and $V_{eff}$ instead of the initial Eqs.~(\ref{Frequoc00})--(\ref{KGoc}). However, if the function $U$ changes the sign, then some solutions can be lost. For example, if $U=U_c$, then we get the dynamical system~(\ref{FOSEQU}). It is easy to see that the point, where $U_c=0$, is not a singular point of this system, whereas the function $P$ and the potential $V_{eff}$  are singular at $U_c=0$.

\begin{figure}[!h]
\centering
\sidecaption
\includegraphics[width=37.2mm]{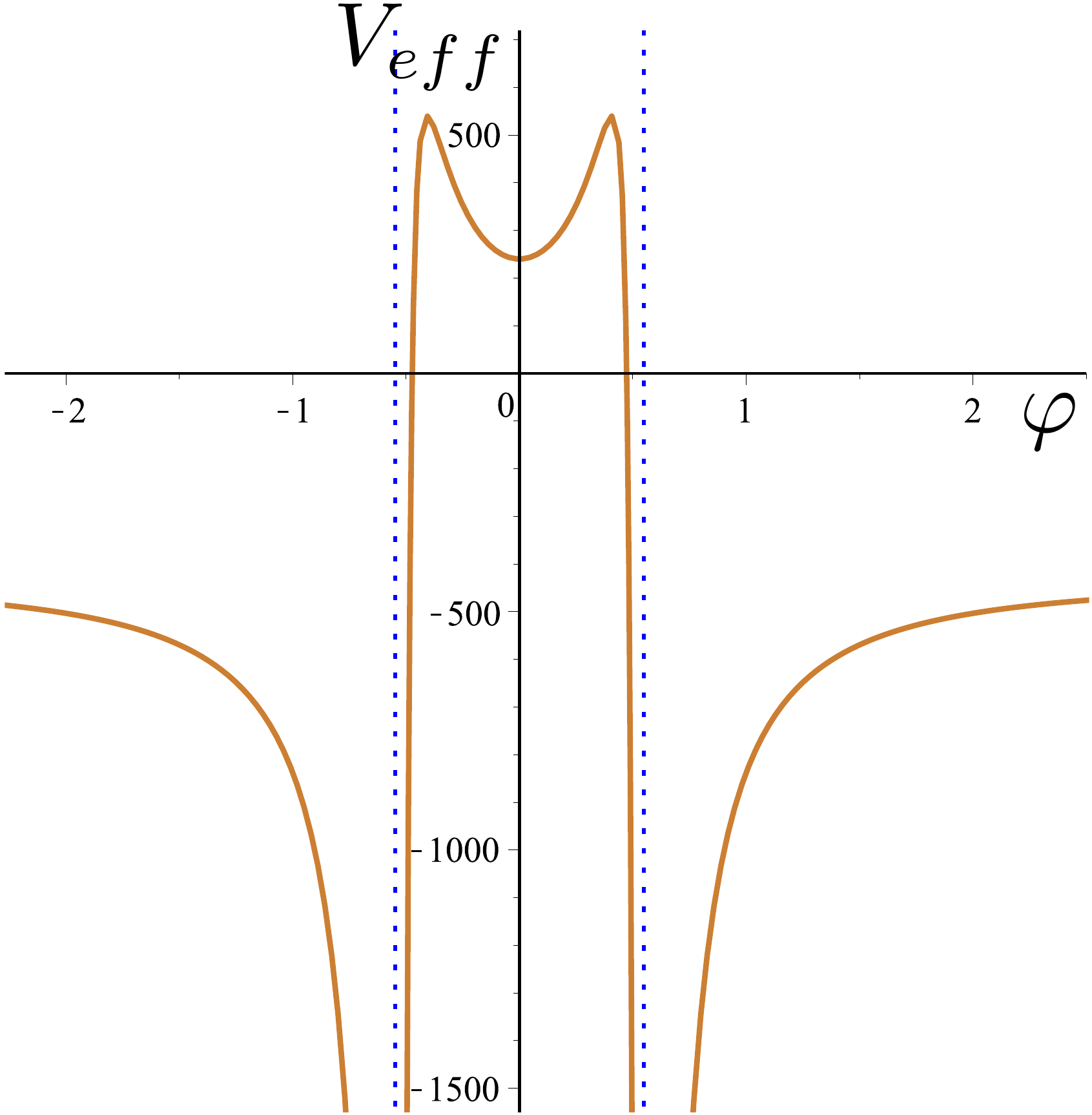}\qquad
\includegraphics[width=37.2mm]{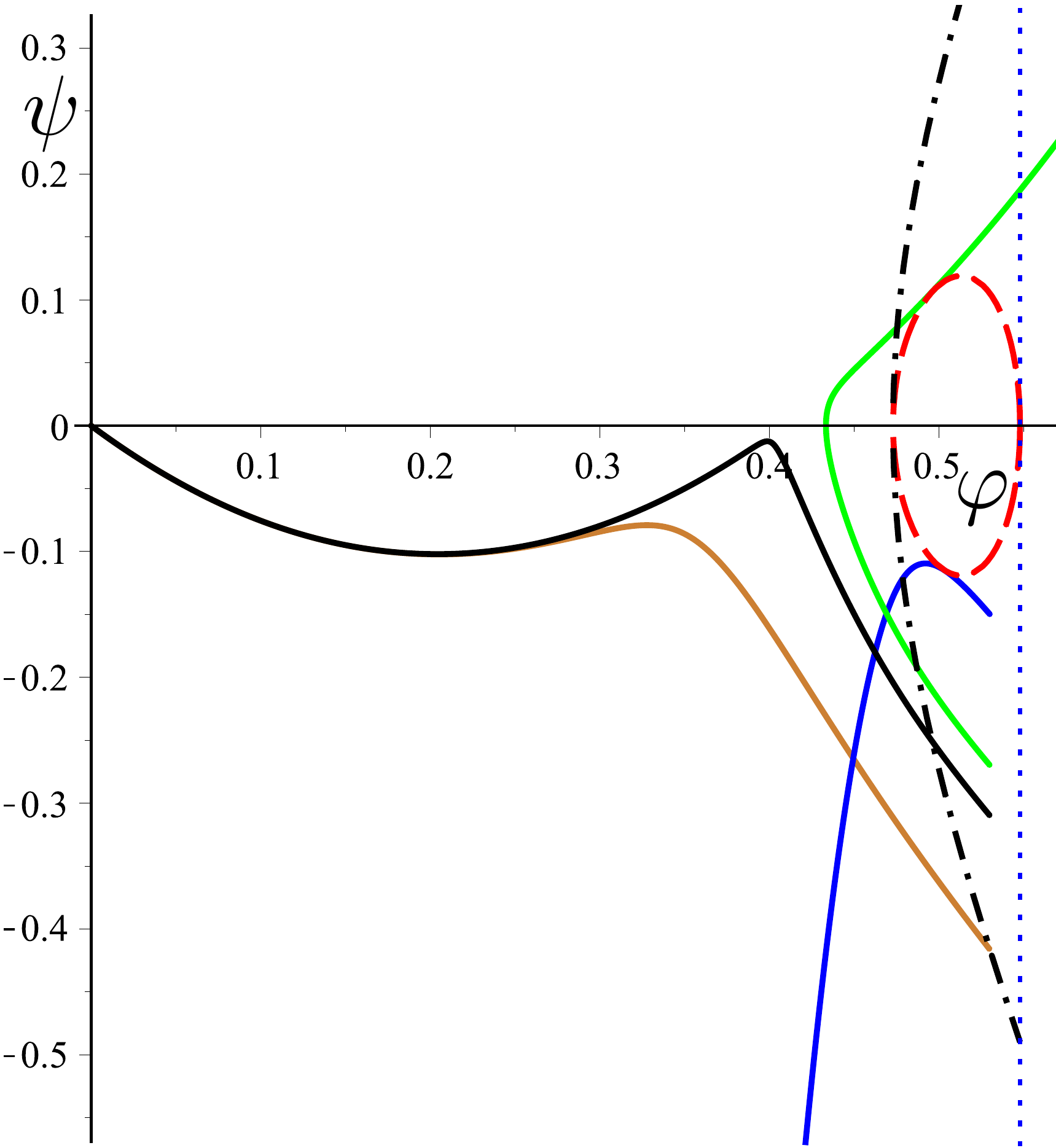}\qquad
\includegraphics[width=37.2mm]{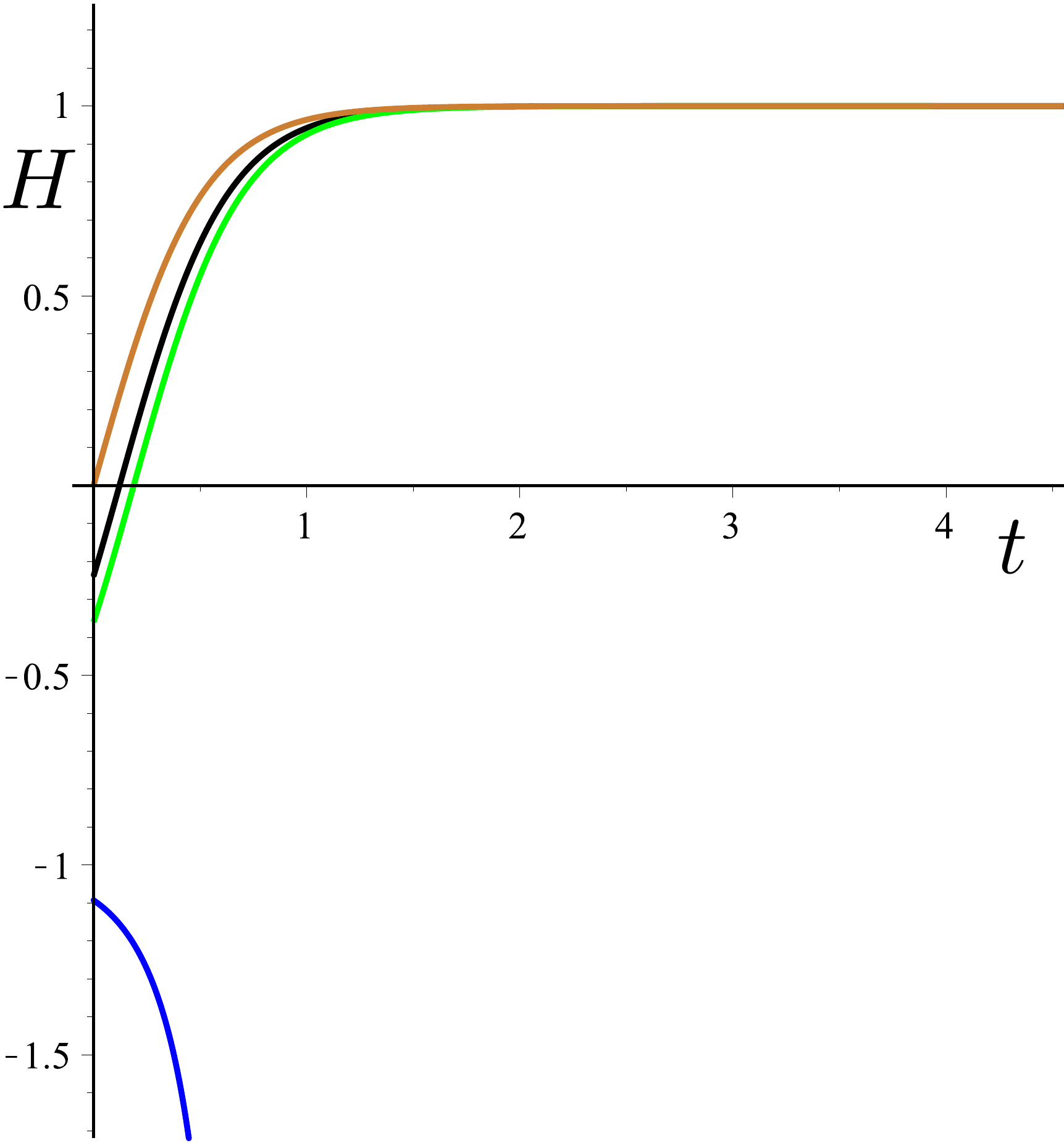}
\caption{The effective potential (left picture), phase trajectories (middle picture) and the Hubble parameter field as function of time (right picture) for $V=C_4\varphi^4+C_0$, $U=U_0-\varphi^2/12$. The parameters are $U_0=1/40$, $C_4=-3$, $C_0=0.15$. The initial values are $\varphi_i=0.53$, and $\psi_i=-0.4164479079$ (gold line),
 $\psi_i=-0.31$ (black line), $\psi_i=-0.27$ (green line), $\psi_i=-0.15$ (blue line),
  The black dash curve corresponds to $H=0$. The blue point lines correspond to $U=0$. The Hubble parameter $H(t)$ is presented in the right picture. }
\label{Casebeta13}
\end{figure}

 In~Fig.~\ref{Casebeta13}  we present the effective potential, phase trajectories and the behavior of the Hubble parameter for the integrable system with a constant $R$. We see that the effective potential has a minimum at $\varphi=0$ and two maxima. The minimum $\varphi=0$ corresponds to a stable de Sitter solution. Any bounce point corresponds to a negative value of the effective potential, so the solution should pass via the maximum  of  $V_{eff}$ to come to zero. The bounce solutions that tends to de Sitter one are denoted by gold and black curves in Fig.~\ref{Casebeta13}. Some solutions does not pass the maximum of  $V_{eff}$  and move to antigravity domain. Note that the Hubble parameter of the bounce solution is finite in this case as well (see green lines in Fig.~\ref{Casebeta13}), but it is not a de Sitter solution. The finiteness of the Hubble parameter that corresponds to infinitely large $\varphi$ is a characteristic property of this integrable model.  The blue curve does not corresponds to de Sitter solution, in this case the initial value of the Hubble parameter is too small and this parameter tends to minus infinity, because $H=H_2(t)$. Colors the Hubble parameter evolutions presented in the right picture of Fig.~\ref{Casebeta13} coincide to the colors of the corresponds phase trajectories in the middle picture.
The behavior of bounce solutions in this model has been studied in detail in~\cite{Boisseau:2015hqa}. In particular in the conformal time ($N=a$)  the function $\varphi(\tau)$ has been found in terms of an elliptic functions.

The red dash curve corresponds to $P=0$. It is the boundary of unreachable domain. Any point inside this curve corresponds to a non-real value of the Hubble parameter, calculated by~(\ref{Frequoc00N1}). Such domain exists at any model with action~(\ref{action}) that has a bounce solution, because  $V(\varphi_b)<0$. The dynamic of solutions of cosmological models with potentials that are not positive definite has been considered in~\cite{Felder:2002jk,Giambo:2014jfa,ABG,ABGV,Giambo:2015tja,PSTV2016}. Solutions can touch the boundary of unreachable domain (see blue curve in~Fig.~\ref{Casebeta13}). The dynamics of such solutions are similar for integrable and non-integrable models and has been described in~\cite{ABGV,PSTV2016}.

\section{Generalizations of the model with a constant $R$}

The considering  model with a constant $R$ belongs to one-parametric set of the integrable cosmological models that has been found in~\cite{KPTVV2015}. To get this set of integrable models and their general solutions the conformal transformation of the metric and the corresponding model in the Einstein frame have been used~\cite{KPTVV2013,KPTVV2015}.

 It was shown in~\cite{Boisseau:2015cda}  that on applying a conformal transformation of the metric $g_{\mu\nu} = \frac{U_0}{U}\tilde{g}_{\mu\nu}$, combined with the use of the new scalar field one arrives to the model with minimally coupled scalar field that has the standard kinetic term and the potential
\begin{equation}
\label{Vint}
    W_c(\phi)  =c_1\cosh^{4}\left(
\frac{\phi}{2\sqrt{3U_0}}\right) + c_2\sinh^{4}\left(
\frac{\phi}{2\sqrt{3U_0}}\right),
\end{equation}
where $c_1$ and $c_2$ are constants and scalar fields $\phi$ and $\varphi$ are connected as follows:
\begin{equation}
 \phi = \sqrt{3U_0}\ln \left[\frac{\sqrt{12U_0}+\varphi}{\sqrt{12U_0}-\varphi}\right]
\qquad\mbox{ and, inversaly, }  \qquad
\varphi = \sqrt{12U_0}\tanh\left[ \frac{\phi}{\sqrt{12U_0}}\right].
\label{connection_c}
\end{equation}

        This minimally coupled model was intensively investigated in~\cite{BarsChen}.
It is known~\cite{Fre} that the model with the potential $W_c$  belongs to the one-parametric set of integrable models with potentials
\begin{equation}
W_\beta(\phi)=c_1\left(\cosh\left[\frac{3\beta\phi}{\sqrt{12U_0}}\right]\right)^{\frac{2(1-\beta)}
{\beta}}+c_2\left(\sinh\left[\frac{3\beta\phi}{\sqrt{12U_0}}\right]\right)^{\frac{2(1-\beta)}{\beta}},
\label{Wgeneral}
\end{equation}
where $\beta$ is an arbitrary constant.
The potential $W_c$ is the potential (\ref{Wgeneral}) at $\beta = 1/3$.

Using the inverse conformal transformation, one can get  from minimally coupled model with $W_\beta$ the integrable model with the function $U_c$ and potential
\begin{equation}
\begin{split}
V_\beta(\varphi)&=\frac{1}{36U_0^2 4^{1/\beta}}\left\{c_1\frac{\left[(\sqrt{12U_0}+\varphi)^{3\beta}
+(\sqrt{12U_0}-\varphi)^{3\beta}\right]^{\frac{2(1-\beta)}{\beta}}}{(12U_0-\varphi^2)^{1-3\beta}}+{}\right.\\ &\left. {}+c_2\frac{\left[(\sqrt{12U_0}+\varphi)^{3\beta}
-(\sqrt{12U_0}-\varphi)^{3\beta}\right]^{\frac{2(1-\beta)}{\beta}}}{(12U_0-\varphi^2)^{1-3\beta}}\right\}.
\end{split}
\label{V-c}
\end{equation}

For an arbitrary $\beta$ the general solution in parametric time has the following form~\cite{KPTVV2015}:
\begin{equation}
\label{GenSolbeta}
\begin{split}
  a&=\frac{1}{2}\left((\vartheta+\eta)^{1/(3\beta)}+(\vartheta-\eta)^{1/(3\beta)}\right), \\
 N&=\frac{2U_0}{3\beta^2}\left((\vartheta+\eta)^{1/(3\beta)}+(\vartheta-\eta)^{1/(3\beta)}\right)
 \left(\vartheta^2-\eta^2\right)^{(1-3\beta)/(3\beta)},\\
 \varphi&=\sqrt{12U_0}\frac{(\vartheta+\eta)^{1/(3\beta)}-(\vartheta-\eta)^{1/(3\beta)}}
{(\vartheta+\eta)^{1/(3\beta)}+(\vartheta-\eta)^{1/(3\beta)}}\, ,
\end{split}
\end{equation}
where $\vartheta$ and $\eta$ are solutions of the following equations:
\begin{equation}
\label{xietaequ}
\dot{\vartheta}^2-8c_1\frac{U_0}{3\beta^2}\vartheta^{\frac{2(1-\beta)}{\beta}}=E,\qquad \dot{\eta}^2+8c_2\frac{U_0}{3\beta^2}\eta^{\frac{2(1-\beta)}{\beta}}=E,
\end{equation}
and $E$ is an arbitrary constant.

For some values of a numeric parameter $\beta$ the potential $V_\beta$ is polynomial. For example, $V_c$ is equal to $V_\beta$ at  $\beta=1/3$.
It is easy to see that at $\beta=1/3$ we get $a=\vartheta$.

In the case $\beta=1$, the potential (\ref{V-c}) can be written as
\begin{equation}
V=V_0\left(U_0-\frac{\varphi^2}{12}\right)^2,
\label{1}
\end{equation}
where $V_0=(c_1+c_2)/U_0^2$.
This potential can be negative if and only if the constant $V_0<0$. Then, on substituting the potential (\ref{1}) into the condition (\ref{bouncecond}), we obtain
$12U_0<\varphi^2$.
So, the model has no bounce solution with $U_c(\varphi_b)>0$.

The third case of a polynomial potential has been proposed in~\cite{KPTVV2015} and corresponds to $\beta=2/3$. In this case
\begin{equation}
\tilde{V}=\frac{c_1}{144U_0^2}\left(12U_0-\varphi^2\right)\left(\varphi^2+2V_1\varphi+12U_0\right),
\label{Vbeta23}
\end{equation}
where $V_1=2c_2\sqrt{3U_0}/c_1$.

Substituting $\beta=2/3$ into (\ref{GenSolbeta}) and (\ref{xietaequ}), we obtain the general solution in the analytic form:
\begin{equation*}
  a=\frac{\sqrt{\vartheta+\eta}+\sqrt{\vartheta-\eta}}{2}, \
 N=\frac{3U_0\left[\sqrt{\vartheta+\eta}+\sqrt{\vartheta-\eta}\right]}{2\sqrt{\vartheta^2-\eta^2}},\  \varphi=\frac{\sqrt{12U_0}\left[\sqrt{\vartheta+\eta}-\sqrt{\vartheta-\eta}\right]}
{\sqrt{\vartheta+\eta}+\sqrt{\vartheta-\eta}}\,,
\end{equation*}
\begin{equation}
\label{xitau}
\vartheta=\frac{1}{6U_0c_1}\left(9U_0^2c_1^2(\tau-\tau_1)^2-E\right),
\qquad \eta={}-\frac{1}{6U_0c_2}\left(9U_0^2c_2^2(\tau-\tau_2)^2-E\right),
\end{equation}
where $\tau_1$, $\tau_2$ and $E$ are arbitrary constants.

For some values of parameters~\cite{KPTVV2015} the model with potential (\ref{Vbeta23}) has a bounce solution with $U_c(\varphi_b)>0$.
We are interesting to a bounce solution that tends to a stable de Sitter ones. Also, we consider only such a solution that $U(\varphi(t))>0$ for all $t\geqslant t_b$.
Let us check whether such a solution exists for the model with $U_c$ and $\tilde{V}$.
To do this we use the effective potential:
\begin{equation}
\label{Veff2}
\tilde{V}_{eff}=\frac{(\varphi^2+2V_1\varphi+12U_0)c_1}{U_0^2(12U_0-\varphi^2)}.
\end{equation}

Its first derivative is equal to zero at the points
\begin{equation}\label{dSphi}
    \varphi_{dS}=\frac{2}{V_1}\left(-6U_0\pm \sqrt{36U_0^2-3U_0V_1^2}\right).
\end{equation}
This result is not valid for $V_1=0$, but in this case $\tilde{V}>0$ at $U_c>0$, so there is no suitable bounce solutions. By this reason we get the first condition on the parameters of potential: $V_1\neq 0$. The second condition is that $ \varphi_{dS}$ should be a real number, so $V_1^2\leqslant 12U_0$. Thus, we~get
\begin{equation}
\label{condV1}
0<V_1^2\leqslant 12U_0.
\end{equation}

A bounce solution corresponds to a negative $\tilde{V}$, whereas de Sitter solution corresponds to a positive $\tilde{V}$, so the necessary condition is that $\tilde{V}=0$ at some points $\varphi_0$ such that
$\varphi_0^2<12U_0$.  The potential $\tilde{V}$ is equal to zero at $\varphi_0=\pm\sqrt{12U_0}$ and
\begin{equation*}
   \varphi_\pm =-V_1\pm\sqrt{V_1^2-12U_0}.
\end{equation*}
So, $\varphi_\pm$ are real only if $V_1^2\geqslant12U_0$. Comparing this condition with (\ref{condV1}) we get $V_1=\pm\sqrt{12U_0}$, but is this case  $\varphi_\pm^2=12U_0$. Thus, we come to conclusion that in the case $\beta=2/3$ no bounce solution tends to de Sitter one.

\section{Concluding remarks}
In this short paper we remind the way to obtain integrable system with non-minimal coupling and compare bounce solutions in two integrable models with polynomial potentials that belong to the one-parameter set of integrable models.

Note that the dynamics of FLRW Universe can be prolonged smoothly into the region with $U<0$ (see, for example~\cite{Skugoreva:2012,KPTVV2016}), however,  anisotropic corrections are expected to diverge when $U$ tends to zero~\cite{Star81,Caputa:2013mfa}. By this reason it is important to analyse the future dynamics of the obtained bounce solutions.

At $\beta=1/3$ there exist bounce solutions that tend to a stable de Sitter solution~\cite{Boisseau:2015hqa}.  At $\beta=2/3$ such solutions do not exist, so in this case bounce solutions tends to antigravity domain, where $U_c<0$.

Note that the monotonically increasing Hubble parameter $H_1(t)$ also is not suitable for construction of a realistic cosmological scenario.
It is possible to get a bounce solutions with non-monotonic behaviour of the Hubble parameter that  tend to de Sitter onesin non-integrable models with slightly modified functions $U_c$ or $V_c$~\cite{Boisseau:2016pfh,PSTV2016}. It would be interesting to construct cosmological model with a non-minimally coupled scalar field, a bounce solution of which is suitable for inflationary scenario.

\textbf{Acknowledgments. }
The authors would  like to thank Alexander~Kamenshchik, Maria~Skugoreva, Alexey~Toporensky, Alessandro~Tronconi, and Giovanni~Venturi for the useful discussions.
Research of E.P. is supported in part by grant MK-7835.2016.2  of the President of Russian Federation.
Research of S.V. is supported in part by grant NSh-7989.2016.2 of the President of Russian Federation.
Researches of E.P. and S.V. are supported in part by the RFBR grant 14-01-00707.

\end{document}